%% file: main.tex
\documentclass[11pt]{article}
\usepackage[T1]{fontenc}
\usepackage[utf8]{inputenc}
\usepackage[margin=1in]{geometry}
\usepackage{lmodern,microtype,amsmath,amssymb,booktabs,tabularx,array,graphicx,enumitem,xcolor}
\usepackage{tikz}

\usetikzlibrary{positioning,arrows.meta}
\usepackage[numbers,sort&compress]{natbib}
\usepackage{xurl}
\usepackage[hidelinks]{hyperref}
\title{Synthetic Cultural Agents from Aggregate Anchors}
\author{Augusto Gonzalez-Bonorino$^{1,2}$ \and Kseniia Biriukova$^{2,3}$ \and Monica Capra$^{2,4}$}
\date{Working paper --- September 2026}
\begin{document}
\maketitle
\begin{center}\small
$^1$Department of Economics, Arizona State University\\
$^2$EconLLM Lab\\
$^3$Department of Information Systems, Arizona State University\\
$^4$Department of Economics, Claremont Graduate University
\end{center}
\input{abstract}
\input{introduction}
\input{litreview}
\input{methodology}
\input{results}
\input{discussion}
\input{conclusion}
\input{disclosures}
\bibliographystyle{unsrtnat}
\bibliography{references}
\clearpage
\input{appendix}
\end{document}

%% file: abstract.tex
\begin{abstract}
Population prompts are widely used to generate synthetic survey responses, but they combine information supplied at inference with associations already encoded during pretraining. We introduce an alternative construction that maps declared aggregate preference anchors into group-indexed choice policies. For each population, the signs of six Global Preferences Survey (GPS) coordinates deterministically label a shared bank of paired synthetic responses, and Direct Preference Optimization fits a parameter-efficient adapter to those comparisons. We evaluate the adapters on candidate World Values Survey (WVS) items using prompts that omit country names and distinguish four questions: recovery of the imposed labels, transfer of the anchor signal to new text, coherence between the GPS anchors and human WVS responses, and agreement between adapter and human scores. The adapters recover the imposed pairwise labels. On a purposively selected sixteen-country development panel, adapter trust scores completely separate the two GPS-sign groups and have a rank correlation of (0.74) with continuous GPS trust scores. Human--GPS and adapter--human associations remain unresolved on the same panel, and results for the other preference dimensions are heterogeneous. These findings show that an anchored policy can retain a declared aggregate signal without thereby reproducing human response patterns. The contribution is therefore both an inspectable construction and an evaluation framework that separates anchor transfer from human criterion agreement.
\end{abstract}


%% file: introduction.tex
\section{Introduction}\label{sec:introduction}
Group-level preference facts are expensive to obtain and easy to over-interpret. A proposed survey item must be piloted, translated, and fielded before researchers can assess whether it measures the intended attribute across populations. Language models offer a cheaper preliminary readout: researchers supply an item and a population description and obtain synthetic responses \cite{argyle2023,horton2023,buskirk2026}. Yet sensitivity to prompting, compressed variance, and unstable subgroup comparisons limit the interpretation of those responses \cite{bisbee2024,boelaert2025,wang2025}. In standard prompting pipelines, population conditioning remains opaque: responses inevitably confound the researcher's intended demographic cues with associations acquired during pretraining. A central methodological question is what population information the researcher explicitly supplies and what subsequent comparisons can establish.

We address this opacity by compiling declared aggregate preference signs into group-indexed parameter adapters. Specifically, we define a synthetic cultural agent as a group-indexed choice policy. It is neither an autonomous agent nor a sample of human respondents. A researcher declares aggregate preference benchmarks, takes their signs to deterministically label a shared bank of contrasting synthetic responses, and fits a parameter adapter for each population using Direct Preference Optimization (DPO) \cite{rafailov2023}. We interpret this loss through a regularized Bradley--Terry pairwise choice model. Each adapter is jointly trained across multiple preference dimensions. At evaluation, the fitted policy assigns probabilities to candidate response codes for new survey items using a country-name-free prompt. Omitting the country name from the evaluation prompt removes that prompting channel. Pretrained associations remain in the base model.

Our empirical application evaluates this construction by linking country scores from the Global Preferences Survey (GPS) \cite{falk2018} to candidate items in World Values Survey (WVS) Wave~7 \cite{wvs2022}. To assess what this pipeline establishes, we separate four distinct evaluation criteria: held-out synthetic-label recovery, cross-text anchor transfer, human cross-survey coherence, and model-human agreement. Internal recovery of synthetic labels confirms that optimization internalizes the training signs, whereas cross-text anchor transfer tests whether the coarse sign transfers to unseen survey items. Human cross-survey coherence evaluates whether human responses on the target instrument align with the external anchor, and model-human agreement measures direct concordance between policy scores and human survey distributions. These readouts target fundamentally different estimands: repeated language-model queries reduce Monte Carlo error around a functional of the model's conditional distribution, whereas larger human samples reduce sampling error around a population functional. Moreover, because the construction relies entirely on published aggregate scores without access to individual-level microdata, the sign vectors do not identify within-group preference distributions. Group-level correspondence across countries faces standard ecological limits, meaning that cross-national alignment cannot establish individual-level measurement validity.

In this application, internal recovery of synthetic labels is strong across all groups on held-out comparisons. On the trust dimension, the fitted policies transfer the coarse GPS sign to new item wording. The comparison with a country-named prompt is descriptive and is reported with the results. However, model-human agreement remains unresolved on this development panel, as does the human survey composite's own coherence with the anchor. Across the remaining preference dimensions, performance diverges: while trust demonstrates directional anchor transfer, other traits display weak or conflicting associations across criteria. Distributional resemblance and directional anchor retention are separate criteria and can favor different saved configurations.

Our contributions are threefold. First, we provide a compiler that maps declared aggregate preference signs into parameter adapters, establishing an inspectable route from population benchmarks to choice policies evaluated without country names. Second, we introduce an evaluation framework that explicitly separates internal label recovery, cross-text anchor transfer, human cross-survey coherence, and model-human agreement. Third, we offer an empirical proof of concept demonstrating that these validation criteria can sharply disagree: a policy can retain an external anchor while agreement with human response distributions remains unresolved. The intended application of this pipeline is group-level analysis and survey pre-piloting as a complement to field research. Whether using such policies for prospective item screening improves item selection before fielding a human survey remains an empirical hypothesis for a later human study.

%% file: litreview.tex
\section{Related research}\label{sec:litreview}
Language models are used both to simulate survey responses and to adapt outputs to cultural communities. These purposes overlap technically, but the information supplied to a model and the evaluation target differ. Prompted synthetic respondents can resemble human samples on selected tasks \cite{argyle2023,horton2023}; subsequent studies document sensitivity, compressed variation, and misportrayal of identity groups \cite{bisbee2024,boelaert2025,wang2025}. Prior synthetic cultural-agent work uses cultural descriptions to study behavior in experiments \cite{capra2025}. We extend this research agenda by making aggregate preference signs the explicit population-specific labeling input.

Prompted simulation supplies a population description at inference and reads associations already present in the model. A structural demand or simulated-method-of-moments exercise would invert a named preference parameter from observed choice probabilities. We compile declared aggregate signs into a scoring policy. The GPS--WVS comparisons report criterion associations for that construction.

Fine-tuning and cultural adaptation already occupy much of this methodological space. Table~\ref{tab:positioning} identifies close comparisons. CultureLLM uses WVS seed data and semantic augmentation for culture-specific fine-tuning \cite{li2024culture}. Cao et al.\ fit country-level survey response distributions and examine cross-survey evaluation \cite{cao2025}. SubPOP learns from subpopulation survey responses and tests transfer to new survey settings \cite{suh2025}.

\begin{table}[htbp]\centering\small
\caption{Selected methodological comparisons of information inputs and evaluation targets.}\label{tab:positioning}
\begin{tabularx}{\linewidth}{@{}>{\raggedright\arraybackslash}p{29mm}>{\raggedright\arraybackslash}X>{\raggedright\arraybackslash}X@{}}\toprule
Approach & Information supplied & Fitting or evaluation target\\\midrule
CultureLLM \cite{li2024culture} & WVS seeds and semantic augmentation & Culture-specific adaptation\\
Cao et al.\ \cite{cao2025} & Country survey distributions & Distributional fit and cross-survey evaluation\\
SubPOP \cite{suh2025} & Subpopulation survey responses & Distribution prediction and transfer\\
Jung and Kim \cite{jung2026} & Korean cultural policy and synthetic triplets & DPO for cultural response coherence\\
Krsteski et al.\ \cite{krsteski2026} & Limited human responses & Allocation to synthesis and rectification\\
Present construction & Aggregate preference signs and a shared contrast bank & Joint adapters; country-name-free external scoring\\\bottomrule
\end{tabularx}\end{table}

Culturally grounded preference optimization is also established in recent work. Jung and Kim generate preference triplets around Korean legal and social norms and fit cultural response policies with DPO \cite{jung2026}. The positive distinction of our approach lies in the specific combination of design choices: an external aggregate-sign input, a shared synthetic bank, a group-specific adapter, a country-name-free evaluation prompt, and an explicit split between anchor transfer and human agreement.

The data requirement differs from direct fitting of a new survey's response distributions: construction uses existing aggregate anchors and synthetic contrasts before human responses to the new instrument are available. That distinction does not imply that scarce human observations should always be spent on training. Krsteski et al.\ examine the roles of prompting, fine-tuning, and statistical rectification when human data are limited \cite{krsteski2026}. Our proposed pre-pilot use must therefore be compared with alternatives that allocate human measurement effort differently.

The econometric interpretation follows work treating language models as conditional response distributions \cite{ludwig2026}, with DPO providing a reference-policy parameterization of Bradley--Terry comparisons \cite{rafailov2023}. The measurement interpretation requires more caution. Construct validity depends on cumulative evidence about the relation between a theoretical attribute and its measures \cite{campbellfiske1959}; cross-national applications also require attention to comparability \cite{steenkamp1998,davidov2014}. Agreement between an anchored policy and a survey may have several explanations, as may disagreement. We therefore report anchor retention and human agreement separately and treat the construction as a complement to human measurement.

%% file: methodology.tex
\section{Methodology}\label{sec:methodology}
We separate the declared population input from the information already present in the models. The researcher chooses aggregate anchors, the synthetic contrast protocol, the fitting procedure, and the evaluation prompt. Declaring these inputs makes the conditioning rule inspectable. It does not restrict all information that can influence the fitted policy.

\subsection{From aggregate anchors to synthetic comparisons}
Let $z_c=(z_{c1},\ldots,z_{cK})$ contain the aggregate scores for group $c$, where $K=6$ coordinates describe trust, risk taking, patience, altruism, positive reciprocity, and negative reciprocity. The fitting protocol uses the binary profile $s_{ck}=1\{z_{ck}\geq0\}$. A teacher and a response generator supply a shared bank $B=\{(x_i,k_i,y_i^+,y_i^-)\}$, where $x_i$ is a scenario and the two responses are intended to express higher and lower levels of dimension $k_i$. These intended loadings are generator judgments, rather than experimentally verified trait manipulations.

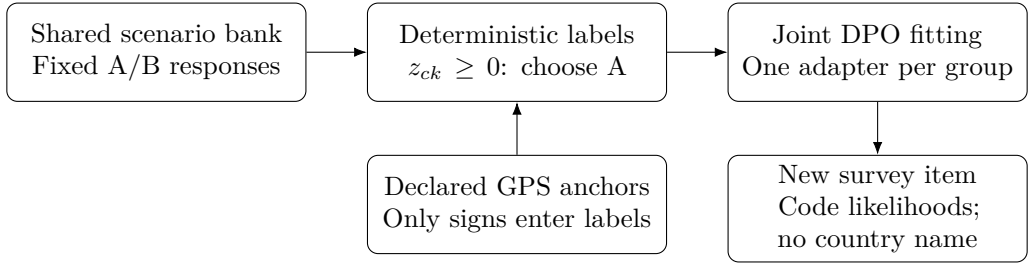
\begin{figure}[htbp]
\centering
\begin{tikzpicture}[>=Latex,node distance=7mm and 8mm,every node/.style={font=\small,align=center},box/.style={draw,rounded corners,text width=37mm,minimum height=13mm}]
\node[box] (bank) {Shared scenario bank\\Fixed A/B responses};
\node[box,right=of bank] (label) {Deterministic labels\\$z_{ck}\geq0$: choose A};
\node[box,right=of label] (fit) {Joint DPO fitting\\One adapter per group};
\node[box,below=of label] (sign) {Declared GPS anchors\\Only signs enter labels};
\node[box,below=of fit] (score) {New survey item\\Code likelihoods; no country name};
\draw[->] (bank)--(label);\draw[->] (sign)--(label);\draw[->] (label)--(fit);\draw[->] (fit)--(score);
\end{tikzpicture}
\caption{Construction and readout. Profile prose and score magnitudes document the anchors but do not feed the deterministic label rule. The shared bank and the pretrained model also influence the resulting policy.}\label{fig:pipeline}
\end{figure}

For each group $c$, the deterministic label assigns the chosen response $y_{w,ci}$ and rejected response $y_{\ell,ci}$ on comparison $i$ as
\begin{equation}
(y_{w,ci},y_{\ell,ci})=\begin{cases}(y_i^+,y_i^-),&z_{c,k_i}\geq0,\\(y_i^-,y_i^+),&z_{c,k_i}<0.\end{cases}
\label{eq:sign_rule}
\end{equation}
The relabeling path reuses fixed scenario and response text. Country identifiers and quantitative profile prose remain metadata; neither is supplied as a country-specific training prompt in this path. Profile rendering documents signs and magnitudes, but magnitudes do not affect these labels. Figure~\ref{fig:pipeline} shows this distinction. Appendix~\ref{app:protocol} gives an actual bank example, prompt excerpts, and source-backed pseudocode.

If two anchor vectors have identical signs, they induce identical labels on the same bank. Holding the split and fitting randomness fixed therefore leaves the fitting problem unchanged. Distinct fitted outputs among such groups cannot be interpreted as recovery of anchor magnitudes. Moreover, a dimension-specific evaluation does not isolate a dimension-specific adapter: all coordinates contribute to the same fitted policy. A changed sign vector defines an intervention on the construction algorithm only. It does not identify a human preference counterfactual.

\begin{quote}\small
\textbf{Source-bank polarity.} The United States is non-negative on all six GPS coordinates, so every training pair is labeled chosen~A. Mexico is negative on all six, so every pair is labeled chosen~B. On this bank, the USA--Mexico contrast is a global polarity flip of the same 658 scenarios.
\end{quote}

The map from declared signs through the shared bank to a fitted policy is a construction. The researcher supplies $z_c$, applies the labeling rule, and obtains the fitted choice policy $\widehat\pi_c$. Identification would require invertibility of a named parameter from observed WVS option probabilities. The present design does not recover $z_c$, the magnitudes $|z_{ck}|$, a single GPS coordinate, or a cultural utility. Rank association between adapter scores and GPS on held-out item text is criterion evidence for the construction. Whether fitted policies differ only as a function of the assigned sign vector remains a separate question.

\subsection{DPO as a pairwise choice estimator}
We treat a language model as a conditional distribution $\pi_\theta(y\mid x)$ over response sequences \cite{ludwig2026}. Under a Bradley--Terry model with latent reward index $r(x,y)$ for response $y$ following prompt $x$,
\begin{equation}
\Pr(y_w\succ y_\ell\mid x)=\sigma\{r(x,y_w)-r(x,y_\ell)\},\qquad \sigma(t)=(1+e^{-t})^{-1}.
\end{equation}
Independent type-I extreme-value utility shocks provide the usual random-utility interpretation \cite{bradleyterry1952}. Here the observed labels are deterministic synthetic choices; we use this likelihood as a fitting criterion rather than as an estimated model of human choice noise.

The KL-regularized policy problem maximizes expected reward minus $\beta D_{\mathrm{KL}}(\pi\Vert\pi_{\mathrm{ref}})$ for $\beta>0$. On the support of the reference policy, its unrestricted optimum $\pi^*$ obeys
\begin{equation}
r(x,y)=\beta\log\frac{\pi^*(y\mid x)}{\pi_{\mathrm{ref}}(y\mid x)}+\beta\log Z(x),
\label{eq:reward}
\end{equation}
where the prompt-specific normalizing term $Z(x)$ cancels from pairwise reward differences. Over the set $\mathcal{D}_c$ of labeled comparisons used to fit group $c$, substitution gives the DPO loss \cite{rafailov2023}:
\begin{align}
\Delta_\theta(x,y_w,y_\ell)&=\log\frac{\pi_\theta(y_w\mid x)}{\pi_{\mathrm{ref}}(y_w\mid x)}-\log\frac{\pi_\theta(y_\ell\mid x)}{\pi_{\mathrm{ref}}(y_\ell\mid x)},\\
\mathcal L_c(\theta)&=-\frac{1}{|\mathcal D_c|}\sum_{(x,y_w,y_\ell)\in\mathcal D_c}\log\sigma\{\beta\Delta_\theta(x,y_w,y_\ell)\}.
\label{eq:dpo}
\end{align}
This is a regularized Bradley--Terry interpretation through the reference-policy parameterization. It does not require an additional explicit KL term in the implemented loss. Finite synthetic comparisons constrain fitted margins on their support; they do not identify all responses, population marginals, or a unique human utility function.

QLoRA restricts the parameters that can move while keeping the quantized base fixed \cite{hu2022,dettmers2023}. We write the fitted policy as $\widehat\pi_c=\pi_{\theta_0+\Delta\theta_c}$, where $\theta_0$ denotes the frozen base parameters and $\Delta\theta_c$ the adapter increment for group $c$. This efficient parameterization changes the feasible policy class, whereas the pairwise likelihood remains the DPO criterion. The unrestricted optimum in equation~\eqref{eq:reward} need not be attained in this class. We retain Llama~3.1 8B Instruct because its limitations in prompted survey simulations provide a relevant setting for testing whether weight-embedded signals survive where prompting is weak \cite{dalloul2026}.

\subsection{Readout and empirical implications}
For an item $q$ presented through rendered chat prompt $x_q$, let $a_{qj}$ be the token sequence of candidate response code $j$, including the prescribed leading space. We compute its teacher-forced log probability by summing conditional token log probabilities,
\begin{equation}
\ell_{cqj}=\sum_t\log\widehat\pi_c(a_{qjt}\mid x_q,a_{qj,<t}),\qquad
\widehat P_c(j\mid x_q)=\frac{\exp\ell_{cqj}}{\sum_h\exp\ell_{cqh}}.
\label{eq:readout}
\end{equation}
The denominator is restricted to the supplied candidate codes. Code tokenization, leading whitespace, item wording, and option order are therefore part of the scoring specification. Evaluating candidate code likelihoods requires an autoregressive model but no sampled answer; optional generated text is a separate output. The normalized code probabilities are a scoring convention whose dispersion need not match population choice frequencies.

For item recode $g_q$, which maps a raw response code $v$ to the value that enters the item score, the policy score is $\sum_jg_q(j)\widehat P_c(j\mid x_q)$. Human scores use the same recode on valid answers before a survey-weighted average. Specifically, Q57 uses the indicator $1\{v=1\}$, Q58--Q64, Q73, and Q81 use $5-v$, Q177 and Q179 use $11-v$, and remaining items keep the raw code. We average item scores within each declared composite and compare country ranks. Appendix~\ref{app:measurement} records the item set, raw-code recodes, missingness rules, and aggregation choices. Group aggregation and aggregation across items are separate operations.

On held-out comparisons, the DPO margin-improvement rate asks whether the fitted relative margin improves over the reference, while pairwise synthetic-label accuracy asks whether the chosen response achieves higher likelihood. Policy--GPS rank association on new text asks whether the coarse anchor sign transfers. Human--GPS association assesses the coherence of a proposed survey composite with the anchor. Policy--human association measures agreement between policy scores and survey country responses. These are distinct empirical implications. Their joint pattern contributes evidence about measurement, but two correlations do not establish construct sharing or measurement invariance \cite{campbellfiske1959,davidov2014,steenkamp1998}.

We assess distributional resemblance separately with
\begin{equation}
\mathrm{TVD}_{cq}=\tfrac12\sum_j|\widehat P_c(j\mid x_q)-P_c^{H}(j\mid q)|,
\end{equation}
where $P_c^{H}(j\mid q)$ denotes the survey-weighted share of valid human answers to item $q$ in group $c$ that equal code $j$. A country-named persona comparator changes the information supplied at inference as well as the conditioning route. The weights-by-prompts comparison is consequently a configuration sensitivity check. Isolating conditioning route would require, among other controls, a prompt receiving the same anchor information. Evaluating the profile adapter with a neutral prompt removes direct country-name conditioning; it does not isolate anchor effects from every other source of variation.

%% file: results.tex
\section{Empirical application and results}\label{sec:results}
Evaluating the fitted policies on the target survey shows coarse trust-sign transfer, while agreement with human respondents remains unresolved. GPS supplies the anchors because its country scores are published aggregates of preference measures selected through prior experimental validation; they are not country-level collections of incentivized experimental choices. WVS Wave~7 serves as the target battery because it is a separate instrument, fielded on overlapping countries, whose items are candidate indicators for those same attributes. The item map is uneven on this panel, which makes the comparison informative: a policy trained on GPS signs can be tested for whether it retains those signs on WVS wording, and the same items can be tested for whether human country order agrees with GPS. Where the candidate map is weak, adapter--human disagreement can arise even when the construction successfully retains the anchor.

Three empirical comparisons structure the findings. First, on trust, a profile adapter with a neutral prompt transfers the coarse GPS sign. The country-prompt comparison is descriptive and appears below. Second, agreement with human country ranks on this panel remains unresolved, as does the candidate map's own association with the anchor across the same sixteen countries. Third, changing the evaluation prompt shifts both rank association and closeness to human response distributions, demonstrating sensitivity to configuration. The exhibits below provide the evidence for these three comparisons.

The diagnostic evaluation uses a purposively selected development panel of sixteen countries whose scores come from two saved jobs: Argentina, Brazil, China, Germany, Egypt, Great Britain, Greece, Indonesia, India, Japan, Mexico, Nigeria, the Netherlands, Russia, Turkey, and the United States. The first wave covers China, Japan, Great Britain, the United States, Mexico, Argentina, Germany, and Russia; the second covers India, Indonesia, Nigeria, Egypt, Turkey, the Netherlands, Brazil, and Greece. We assembled those waves to include both global sign polarities, the three duplicated sign profiles that share a labeling problem, mixed profiles, and geographic spread. The panel is not a probability sample of the forty-two-country GPS--WVS intersection. A later study that treats country selection as a sampling problem should train every country in that intersection or a predeclared random subset of it.

The primary comparison holds countries, item sets, recodes, and aggregation rules fixed across arms. It contains twenty-three single-choice items, including nine trust items. Egypt has no valid Q69--Q71 observations and Great Britain has no valid Q174 observations, so these items are removed everywhere. Q12--Q14 are multiple-response items and are excluded from the categorical softmax evaluation. Table~\ref{tab:map} lists the resulting map. Its proposed composites should not be read as validated measures of all six GPS traits. The original mapping preceded model scoring, whereas the strict common-support analysis and its sensitivity checks are retrospective.

\begin{table}[htbp]\centering\small
\caption{Candidate indicators on the strict common-support panel. Each row defines a candidate item set; latent-scale validity remains to be assessed. Patience items are interpreted separately.}\label{tab:map}
\begin{tabularx}{\linewidth}{@{}>{\raggedright\arraybackslash}p{30mm}>{\raggedright\arraybackslash}p{35mm}>{\raggedright\arraybackslash}X@{}}\toprule
Anchor & WVS items & Interpretation to be assessed\\\midrule
Trust & Q57--Q64, Q73 & Generalized, interpersonal, and institutional trust/confidence\\
Patience & Q43; Q50 & Less importance placed on work; financial satisfaction\\
Risk taking & Q106, Q107, Q109, Q178 & Economic values and fare avoidance\\
Positive reciprocity & Q81 & Confidence in charitable organizations\\
Negative reciprocity & Q176, Q177, Q179, Q195 & Moral clarity and justifiability judgments\\
Altruism & Q99, Q101, Q103 & Membership in environmental, charitable, and self-help groups\\\bottomrule
\end{tabularx}\end{table}

Generalized trust exhibits weak human--GPS agreement on the sixteen-country panel ($\rho = 0.15$ on Q57), underscoring that equivalence between each listed WVS item and its proposed GPS measure remains to be assessed. Q57 is the generalized-trust item related to the GPS validation comparison \cite{falk2018}. The wider trust composite pools substantively different items. The remaining mappings are exploratory bridges, especially where Wave~7 lacks direct counterparts to the preference benchmark.

\subsection{Recovery of the synthetic labels}
The country adapters instantiate thirteen distinct six-sign profiles across the sixteen countries (Table~\ref{tab:signprofiles}): Mexico shares its signs with Russia, India with Greece, and Indonesia with Egypt. Thus there are fewer distinct labeling problems than country labels. Each country adapter uses 526 training comparisons from a shared 658-pair bank, with 132 held-out comparisons. The fitting configuration uses $\beta=0.1$, LoRA rank 16 and scaling parameter 32, and one epoch. Appendix~\ref{app:training} records configuration and provenance limits.
\input{tables/sign_profiles}

Held-out evaluations confirm that the adapters learned the imposed synthetic contrasts. Pair-pooled across 2,112 held-out country--pair cells, the DPO margin-improvement rate is $0.942$ and pairwise synthetic-label accuracy is $0.819$. The former records a positive improvement in the chosen--rejected margin over the reference; the latter records a positive adapted-model margin itself. Country-level DPO margin-improvement rate ranges from $0.886$ to $0.992$. These results support recovery of imposed synthetic contrasts, but they do not measure prediction of independently observed choices. The inspected training notebooks record a shared seed-42 split of 526 and 132, but they do not record why Russia's holdout differs from that split. Fifteen countries share the seed-42 holdout; Russia's different 132-pair holdout shares 26 prompts with that set. This discrepancy limits the interpretation of pooled precision and country comparisons.

\subsection{Cross-survey associations}
\begin{figure}[htbp]\centering
\includegraphics[width=\linewidth]{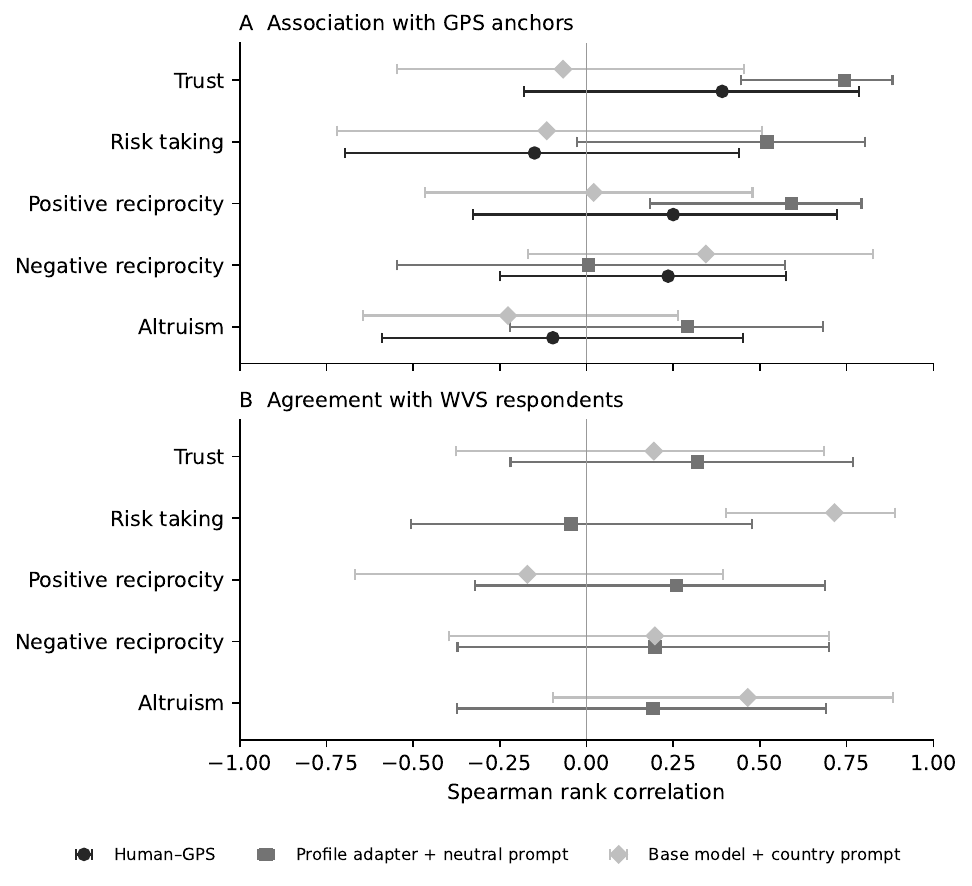}
\caption{Sixteen-country panel. Panel A is association with the GPS anchors. Panel B is agreement with WVS country responses. Markers are Spearman correlations; bars are 95\% country-bootstrap intervals from 2,000 resamples. The intervals are conditional within-panel stability summaries. They exclude training-seed variation, respondent resampling, and GPS measurement error. Patience is omitted because Q43 and Q50 are reported separately.}\label{fig:associations}
\end{figure}

Country-rank associations indicate marked differences between anchor retention and alignment with human respondents (Figure~\ref{fig:associations}). Table~\ref{tab:persona} supplies the country-named persona--GPS comparator. Country-bootstrap intervals use 2,000 common resamples and describe conditional within-panel stability given the fitted policies, GPS anchors, WVS point estimates, and selected items. They do not represent population sampling, respondent uncertainty, GPS measurement error, or training-seed variation, and are not multiplicity-adjusted. The intervals are wide because each association evaluates rank correlation across sixteen countries.

On trust, profile adapters evaluated with a neutral prompt transfer the coarse GPS sign, with rank correlation $\rho = 0.74$ $[0.44, 0.88]$, compared with $-0.07$ for the country-prompted persona arm. The paired adapter-minus-persona difference is $0.81$ $[0.16, 1.31]$. The persona--GPS comparator is descriptive, with the caveat that identical checkpoints, tokenizers, code strings, and scoring procedures across historical runs were not established. Human--GPS association on the sixteen-country rectangle is $0.39$ $[-0.18, 0.79]$, whereas adapter--human association is $0.32$ $[-0.22, 0.77]$. The sixteen-country intervals leave human agreement unresolved, and they also leave the map's own criterion association unresolved. On the same nine-item recodes, the forty-two-country GPS--WVS intersection yields human--GPS association $0.35$ $[0.09, 0.56]$ (Table~\ref{tab:human42}). That larger human panel provides a criterion check on the map, but the adapters were not retrained on it. We draw no inference from similarity between a correlation and a product of other correlations, as that comparison does not identify an underlying mechanism.

A sign check on the trust composites establishes coarse trust-sign transfer across the purposively selected development panel. Seven countries have non-negative GPS trust anchors and nine have negative anchors. Adapter trust scores completely separate these two groups (rank-biserial correlation = 1.00) \footnote{A non-parametric measure that assesses the strength and direction of association between one binary (dichotomous) variable and one ordinal (ranked) variable}, with a mean between-group score difference of 0.43. We describe this result as coarse trust-sign transfer. The sixteen scores come from two saved jobs rather than a single evaluation run. The reported bootstrap intervals are conditional within-panel stability summaries; they do not reflect population sampling, respondent uncertainty, GPS measurement error, or training-seed variation.

In the tables, $\rho$ denotes the Spearman rank correlation across countries.
\input{tables/persona}

Across the remaining preference dimensions, anchor retention and human agreement vary substantially by trait. For risk taking, adapter--GPS association on the two-wave scores is $0.52$ $[-0.03, 0.80]$, while adapter--human association is $-0.04$. Positive reciprocity yields an adapter--GPS association of $0.59$ and adapter--human association of $0.26$ on its single item (Q81). Altruism yields an adapter--GPS association of $0.29$ and adapter--human association of $0.19$. Negative reciprocity displays an adapter--GPS association of $0.01$ and adapter--human association of $0.20$. These divergences illustrate that anchor retention on one construct does not imply uniform transfer across all constructs. Mapping choices, anchor measurement, ecological differences, bank semantics, and scoring limitations can all contribute to divergence across dimensions.

\input{tables/patience}
The two patience candidate indicators exhibit contrasting empirical patterns (Table~\ref{tab:patience}). On the two-wave scores, Q43 yields an adapter--GPS association of $0.01$ and adapter--human association of $0.29$, whereas Q50 yields an adapter--GPS association of $0.44$ and adapter--human association of $0.25$. The human--GPS correlations similarly diverge in sign: Q43 concerns reduced importance of work, whereas Q50 concerns financial satisfaction. We retain their original coding and report them separately rather than selecting a polarity after observing the correlations. Their two-item mean is an available computational summary, but it is not treated as a validated patience measure. The primary analysis therefore avoids aggregating them into a single composite.

\subsection{Distributional fit and prompt sensitivity}
Directional anchor retention and closeness to human response distributions do not favor the same configuration (Table~\ref{tab:twobytwo}). Evaluating the profile adapter with a neutral prompt achieves a trust--GPS association of $0.74$ and a mean total variation distance (TVD) of $0.487$ across 368 country--item cells. Adding a country prompt yields a trust--GPS association of $0.67$ and a mean TVD of $0.401$, though that trust cell was not recomputed from the two saved jobs. The base model with a country prompt achieves the lowest TVD ($0.369$), but exhibits no positive rank association with the anchor ($\rho = -0.07$).

\begin{table}[htbp]\centering\small
\caption{Configuration sensitivity on the strict panel. TVD averages 368 country--item cells. The Profile adapter + country prompt trust cell ($0.67$) was not recomputed from the two saved jobs. The base TVD uses banked runs. No base rank association with the anchor is interpreted, because run variation is confounded with the bank. Rank associations on this table use the same country bootstrap as Figure~\ref{fig:associations}, which excludes training-seed variation and GPS measurement error.}\label{tab:twobytwo}
\begin{tabular}{lcc}\toprule
Configuration & Trust--GPS $\rho$ & Mean TVD\\\midrule
Base model + neutral prompt & --- & 0.462\\
Base model + country prompt & $-0.07$ & 0.369\\
Profile adapter + neutral prompt & 0.74 & 0.487\\
Profile adapter + country prompt & 0.67 & 0.401\\\bottomrule
\end{tabular}\end{table}

Configuration sensitivity is trait-dependent, demonstrating that adding a country-specific prompt shifts both rank association and distributional fit. On risk taking, adding a country prompt alters the direction of association with the anchor and shifts alignment with human responses. These shifts describe configuration sensitivity rather than separating latent priors from anchor effects or establishing a general division between weights determining direction and prompts determining shape. The historical base runs also vary numerically despite identical saved prompts because they are not checkpoint-equivalent; they are therefore not treated as a single fixed base distribution. Identical checkpoints, tokenizers, code strings, and scoring procedures were not established across those runs, leaving comparisons with the base model descriptive.

The empirical evidence supports a fitted synthetic choice policy with trait-dependent external anchor association. Reliable prediction of human country ranks and practical pre-pilot benefits remain unestablished. We do not mix thirty-item total variation distance levels from earlier exploratory diagnostics with the strict-panel estimates, omitting that exhibit pending reconciliation of item eligibility and historical aggregation. A rank association alone cannot establish population dispersion.

%% file: tables/sign_profiles.tex
\begin{table}[htbp]\centering\small
\caption{Thirteen distinct GPS sign profiles among the sixteen adapter countries. Coordinates are ordered trust, risk taking, patience, altruism, positive reciprocity, and negative reciprocity, matching Appendix~\ref{app:protocol}. The United States is non-negative on every coordinate; Mexico and Russia are negative on every coordinate. Identical sign vectors induce identical labels on the shared bank.}\label{tab:signprofiles}
\begin{tabular}{lcccccc}\toprule
Members & trust & risk & patience & altruism & pos.\ recip. & neg.\ recip. \\\midrule
EGY, IDN & $+$ & $-$ & $-$ & $+$ & $+$ & $+$ \\
GRC, IND & $-$ & $-$ & $-$ & $-$ & $-$ & $+$ \\
MEX, RUS & $-$ & $-$ & $-$ & $-$ & $-$ & $-$ \\
ARG & $-$ & $+$ & $-$ & $+$ & $+$ & $-$ \\
BRA & $-$ & $-$ & $-$ & $+$ & $+$ & $-$ \\
CHN & $+$ & $-$ & $+$ & $+$ & $+$ & $+$ \\
DEU & $-$ & $-$ & $+$ & $-$ & $-$ & $-$ \\
GBR & $+$ & $+$ & $+$ & $+$ & $-$ & $+$ \\
JPN & $-$ & $-$ & $+$ & $-$ & $-$ & $+$ \\
NGA & $-$ & $+$ & $-$ & $-$ & $-$ & $+$ \\
NLD & $+$ & $+$ & $+$ & $-$ & $-$ & $+$ \\
TUR & $+$ & $+$ & $-$ & $-$ & $-$ & $+$ \\
USA & $+$ & $+$ & $+$ & $+$ & $+$ & $+$ \\
\bottomrule
\end{tabular}\end{table}

%% file: tables/persona.tex
\begin{table}[htbp]\centering\small
\caption{Country-named persona--GPS association on the strict panel, with 95\% country-bootstrap intervals. This supplies the anchor comparator for Figure~\ref{fig:associations}.}\label{tab:persona}
\begin{tabular}{lcc}\toprule Anchor & $\rho$ & Interval\\\midrule
Trust & -0.07 & [-0.55, 0.45] \\
Risk taking & -0.11 & [-0.72, 0.51] \\
Positive reciprocity & 0.02 & [-0.47, 0.48] \\
Negative reciprocity & 0.34 & [-0.17, 0.82] \\
Altruism & -0.23 & [-0.64, 0.26] \\
\bottomrule
\end{tabular}\end{table}

%% file: tables/patience.tex
\begin{table}[htbp]\centering\small
\caption{Patience candidate indicators, separately scored on sixteen countries. Entries are Spearman correlations; brackets give 95\% country-bootstrap intervals.}\label{tab:patience}
\begin{tabular}{lcc}\toprule Pairing & Q43 & Q50\\\midrule
Human--GPS & -0.56 [-0.86, -0.04] & 0.62 [0.11, 0.91] \\
Adapter--GPS & 0.01 [-0.47, 0.44] & 0.44 [-0.12, 0.79] \\
Adapter--human & 0.29 [-0.19, 0.73] & 0.25 [-0.32, 0.70] \\
Persona--GPS & -0.62 [-0.90, -0.10] & 0.62 [0.16, 0.87] \\
Persona--human & 0.59 [0.11, 0.82] & 0.70 [0.23, 0.91] \\
\bottomrule
\end{tabular}\end{table}

%% file: discussion.tex
\section{Discussion}\label{sec:discussion}
The construction makes the rule translating declared aggregate preference signs into training labels explicit and auditable. A policy fitted to those labels can score new survey text without naming the population in the evaluation prompt, providing a controlled computational object for examining how supplied group information affects synthetic responses. This restriction applies to the declared conditioning channel, whereas pretrained representations and generator bank semantics remain embedded in the fitted model. We intend these policies to complement field work rather than replace human subjects or ascribe fixed cultural essences.

The empirical application demonstrates that anchor fidelity and human resemblance are distinct. Recovery of synthetic labels confirms that optimization moved pairwise contrasts in the intended direction, and the fitted policies transfer the coarse trust sign to unseen survey items. Distributional resemblance to human choices is reported separately in Section~\ref{sec:results}. That comparison describes two saved configurations. Identical checkpoints, tokenizers, code strings, and scoring procedures were not established. Agreement with human country responses remains uncertain on this panel, while exploratory dimensions reveal substantial cross-survey divergence. The fitted policies are scoring instruments; construct validity requires independent human evidence.

A prospective application of this construction is survey pre-piloting, where researchers score candidate questionnaire items against declared benchmarks to nominate items before fielding. However, our results do not demonstrate that synthetic scoring reduces fielding costs, improves item selection, or preserves validity in practice. Prospective pre-pilot item screening still needs an independent human study. Specifically, evaluating this workflow requires a blinded comparison of adapter-assisted selection against expert selection on a held-out human pilot, following the validation design in Appendix~\ref{app:humanscreen}. Scoring candidate items with synthetic respondents cannot establish measurement validity, and empirical human data remain indispensable for calibrating and rectifying model-assisted measures \cite{krsteski2026}.

Several limitations bound this evidence. The country panel is a purposively selected development panel of sixteen economies rather than a probability sample of the joint survey intersection. Duplicated sign profiles mean that countries with the same six signs share the same training labels, and sign-only supervision discards anchor magnitudes. In addition, the scenarios rely on a single shared bank whose loadings reflect generator judgments rather than experimentally verified trait manipulations. Omitting country tokens at evaluation does not remove pretraining exposure to public surveys, and joint optimization across traits prevents attributing item responses exclusively to any single anchor dimension.

%% file: conclusion.tex
\section{Conclusion}\label{sec:conclusion}

We construct synthetic cultural agents by compiling the signs of declared aggregate preference anchors into group-indexed choice policies. A common bank of contrasting responses is labeled according to each group's six-dimensional GPS sign profile, and a parameter-efficient adapter is fitted to those comparisons. Evaluating the resulting policies without country names in the prompt makes this population-specific labeling channel explicit and auditable. The fitted policy nevertheless remains jointly shaped by the pretrained model, the synthetic bank, and the complete sign profile.

The empirical application shows why anchor transfer and human agreement must be evaluated separately. The adapters learned the imposed synthetic comparisons, and their trust scores transferred the coarse GPS sign to new WVS wording. This pattern did not generalize uniformly across the remaining preference dimensions. Moreover, agreement with human country responses remained unresolved on the sixteen-country development panel. The central finding is therefore not that the adapters reproduce national cultures. It is that retaining a declared anchor and resembling human respondents are distinct empirical achievements: success on the former does not establish the latter.

The appropriate interpretation of these adapters is consequently as inspectable scoring instruments rather than synthetic populations. They provide a way to study whether a declared aggregate signal survives transport to new item text, while independent human evidence remains necessary to determine whether that signal corresponds to the intended construct. The next test is prospective: predeclare a set of candidate survey items, compare adapter-assisted and expert item selection, and evaluate both against a blinded human pilot. Such a study would determine whether anchored scoring improves survey pre-piloting rather than merely reproducing its inputs.

%% file: disclosures.tex
\section*{Code, data, and disclosures}
The repository is \url{https://github.com/EconLLM-Lab/SCA2_PofW}. The reported correlations are reprinted from the committed country--item score file in that repository. The zip linked here is an earlier archive of model outputs from 17 September 2026 and is not the source of the reported trust correlation of 0.74. It contains no survey microdata, at \url{https://drive.google.com/uc?export=download&id=1lIAx0ueSpgaZPmAzNbFSD31ddGQH7Nqo}. Adapter weights are released at \url{https://huggingface.co/Bonorinoa/SCA2-phase2-adapters}, commit \texttt{03c43dfd27c9535283feb78ffa727da3bfa1b966}. A Hugging Face account and acceptance of the repository gate are required to download them. Reprinting the reported tables uses the committed score file. Scoring new item text uses the adapters.

GPS and WVS data remain subject to their providers' access and use conditions \cite{falk2018,wvs2022}. The package does not redistribute respondent microdata. The analysis is at the country level and should not be used to infer an individual's preferences or to portray national populations as homogeneous. Independent human work would require its own consent and ethics procedures.

The study was not preregistered. Common-support corrections, the forty-two-country human map, and reported sensitivity checks are retrospective. The authors declare no funding and no competing interests. Language models were used for writing and coding assistance. The authors remain responsible for the claims, analysis, and final manuscript.

%% file: appendix.tex
\appendix
\section{Construction protocol and worked example}\label{app:protocol}

The implemented manifest identifies the sign relabeling run as \path{gps_sign_relabel_all_20260816_004545Z}. Its source is the June 2026 USA--Mexico fixed-response bank. Relabeling deduplicates by dimension and prompt, rejects conflicting A/B text, and applies the sign rule to the recovered common bank. The exported countries exceed the trained panel; an exported label file does not imply that an adapter was fitted. The generation manifest sets \texttt{use\_anchors=false}, so optional WVS anchors are not an input to this recorded source run.

The following pseudocode describes the implemented logic in \path{synthetic_generation/sca2_datagen/relabel.py}, especially \texttt{build\_triplet\_bank}, \texttt{sign\_choice}, and \path{label_triplet_for_country}. It is explanatory pseudocode rather than a claim that the complete replication package has been released.
\begin{quote}\ttfamily\small
Recover shared (scenario, dimension, response A, response B) bank.\\
Deduplicate on dimension and scenario; fail on conflicting A/B text.\\
For each group and each shared pair:\\
\hspace*{1em}Read finite GPS score for the pair's target dimension.\\
\hspace*{1em}Choose A when score >= 0; otherwise choose B.\\
\hspace*{1em}Export scenario, chosen text, rejected text; retain metadata.\\
Fit one adapter jointly on the group's training comparisons.\\
Score candidate codes for new items with the fixed evaluation prompt.
\end{quote}

\subsection{A real fixed-response contrast}
The example below is triplet \texttt{00023ac74dac} in \texttt{triplets\_bank.jsonl}. The target is positive reciprocity. Argentina's exported score is $+0.1596793085$; its chosen option is A. This is an example of a deterministic training label. It does not describe an observed Argentine choice.
\begin{quote}\small
\textbf{Scenario.} A fellow student tutored me for free before a critical exam, and now they're facing a tight deadline for their own project and need someone to review their work. Decision: I can either take the time to thoroughly review their project or concentrate on my own study schedule. Trade-off: Helping my fellow student would acknowledge the effort they put into tutoring me, but it would cut into my own study time and potentially impact my performance.

\textbf{Response A.} I'll set aside time to carefully review their work this evening. They went out of their way to help me when it mattered most, and I want to make sure they know their effort made a difference. If I don't take this seriously, it would feel like I'm ignoring the support that helped me succeed.

\textbf{Response B.} I'll let them know I can only do a quick scan tomorrow morning. My exam prep schedule is already packed with material I haven't mastered yet, and I need to be realistic about what I can manage without burning out. I hope they understand my limitations in this situation.
\end{quote}
The Argentina metadata records signs $(-,+,-,+,+,-)$ in the order trust, risk taking, patience, altruism, positive reciprocity, negative reciprocity. Its policy is fitted jointly across these signs. A new WVS item, such as Q81 on confidence in charitable organizations, is then scored using the fitted adapter and the fixed code-likelihood readout. The scenario and Q81 are different texts. Their shared proposed dimension is an assumption of the map, and is not established by this example.

\subsection{Profile documentation and generation prompts}
The rendering function \texttt{build\_cultural\_profile} in \texttt{profiles.py} starts with ``GPS CULTURAL STATE VECTOR (Falk et al. 2018, standardized deviations from global mean):''. Each line has the format ``symbol (dimension) = signed score: magnitude. description''. It describes $|z|<0.10$ as ``near the global average'', $0.10\leq|z|<0.40$ as ``moderately above/below average'', and larger values as ``strongly above/below average''. Scores are printed to two decimals. These thresholds describe profile prose; they do not modify the implemented sign labels. A legacy model-selection path used profile prose, but it is distinct from the deterministic path examined here.

The current generation source records the following exact scenario-template instructions:
\begin{quote}\small
Context: One realistic sentence establishing the agent, setting, and stakes.\\
Decision: One sentence stating the two behaviorally plausible options the agent is choosing between.\\
Trade-off: One sentence making the core target-facet tension explicit without naming high/low GPS scores.
\end{quote}
The fixed-pair prompt in \texttt{generate.py} includes these exact requirements:
\begin{quote}\small
Generate two opposing responses to this same scenario.\\
- Response A should load positively on the target dimension: it should express a higher level of the target trait/facet.\\
- Response B should load negatively on the target dimension: it should express a lower level or absence of the target trait/facet.\\
- Positive loading does not mean morally better, more polite, or more socially desirable.\\
- Negative loading does not mean irrational, careless, hostile, or cartoonishly selfish.
\end{quote}
Other instructions request matching perspective, tone, length, stakes, and the remaining traits, and prohibit country labels. They express design intentions rather than verified controls. Complete source prompt templates are transcribed in the accompanying \texttt{PROMPT\_SOURCE\_EXCERPTS.md}. Current source text is distinguished there from archived realized evaluation prompts; a complete historical request log tying every generation call to an immutable provider revision has not been verified.

The source manifest names endpoint aliases \texttt{hf-teacher}, \texttt{hf-generator}, and \texttt{hf-scorer}, with temperatures 0.7, 0.8, and 0.1. Current configuration maps them to Llama-3.3-70B-Instruct, Qwen3-32B, and Phi-4 endpoint names. Immutable served revisions remain unverified. The sixteen adapters examined here were produced by deterministic sign relabeling of the shared bank; the scorer was not an implementation gate. The manifest's requested scenario count does not by itself explain the size of the recovered bank; we use the materialized bank and its 658 unique pairs as the recorded bank.

\subsection{Evaluation prompt and code scoring}
The saved evaluation preview and \texttt{build\_survey\_user\_prompt} agree on the prompt wrapper without country tokens:
\begin{quote}\small
Answer this questionnaire as an individual person. Respond naturally and sincerely, as someone would in real life. Do not mention being an AI or assistant.

Question:\\
\emph{[item text]}

Select exactly one response code from the options below.\\
\emph{[code: label, one option per line]}

Return only the response code, with no words or explanation.

Answer:
\end{quote}
For the persona condition, the first sentence becomes ``Answer this questionnaire as a typical adult living in \emph{[country name]}.'' The tokenizer applies the chat template with a generation prompt. Candidate completion strings are a leading space followed by a code. Their log probabilities are summed without length normalization, then normalized across codes at unit temperature. This is the readout in equation~\eqref{eq:readout}. Saved prompt previews include the rendered chat wrapper. The complete realized wrapper for Q57 is reproduced in the accompanying prompt document; it should be preserved when replaying the stored specification.

\section{Training configuration and historical diagnostics}\label{app:training}
The inspected CO2 multi-country notebook records Llama-3.1-8B-Instruct, 4-bit NF4 with double quantization and fp16 base compute, attention-projection LoRA modules \texttt{q\_proj}, \texttt{k\_proj}, \texttt{v\_proj}, and \texttt{o\_proj}, rank 16, scaling 32, and dropout 0.05. It uses DPO beta 0.1, learning rate $10^{-4}$ with cosine scheduling, one epoch, per-device batch size 1, and 16 gradient-accumulation steps. The maximum length is 768 with \texttt{keep\_end} truncation. Trainable parameters are converted to fp32 after trainer initialization. The source has the warmup-ratio assignment commented out, so the prior draft's claim that every adapter used warmup 0.03 is not retained.

These are inspected notebook settings. Their correspondence with every stored checkpoint remains to be established. The replication package still needs an immutable checkpoint-to-run configuration manifest, exact dependency versions, and reference-log-probability provenance. Stored evidence supports the shared-bank 526/132 training/holdout counts and the Russia split exception reported in the main text. The inspected notebooks do not record why Russia's holdout differs from that shared split. The notebook records a Tesla T4 run; account tier and historical billing have not been established, so no free-compute or cost claim is made.

A historical language-model diagnostic scored generated responses on all six GPS dimensions. It did not exclude pairs from implemented sign relabeling. Pair loadings remain generator judgments, and the shared bank is mixed-trait. Unused process diagnostics are not evidence of trait isolation. A new bank or a newly imposed exclusion rule would be a separate construction experiment.

\section{Measurement specification and sensitivity status}\label{app:measurement}
The strict panel uses Table~\ref{tab:map}'s common item set for every country and arm. We remove raw WVS missing codes $-1$ through $-5$ and nonfinite observations before any transformation, retaining valid zero membership responses. Human item means use positive \texttt{W\_WEIGHT} weights and require at least fifty valid positive-weight responses per cell. All retained strict-panel cells pass that requirement in the saved audit.

The recodes inherited from the declared analysis are $g_{57}(v)=1\{v=1\}$; $g_q(v)=5-v$ for Q58--Q64, Q73, and Q81; and $g_q(v)=11-v$ for Q177 and Q179. Other retained items use their raw codes. For a model, each code is recoded before taking its probability-weighted expectation. For humans, each valid answer is recoded before its weighted mean. Proposed dimension composites are unweighted averages of these item means, without standardizing their ranges. This makes scale ranges part of the composite definition. Only the trust composite mixes item ranges. Mapping those items onto a common unit interval before averaging does not change the sign of the reported trust association or whether its interval excludes zero. The reported figure remains the raw-scale correlation of 0.74. Other reported composites use one range within the dimension, so that rescaling leaves their rank correlations unchanged.

An earlier path transformed negative missing codes before exclusion. The corrected analyses exclude missing values before recoding, and Q69--Q71 enter only the sensitivity surface. A corrected broader human-only comparison uses a different country sample and must not be substituted for a matched model--human comparison. We retain Egypt in the primary sample. Earlier numeric Egypt robustness checks were computed on a superseded score file and are not reported as checks of the current trust result.

Table~\ref{tab:items} details the 26 candidate items, their proposed GPS links, recode transformations, valid ranges, and analysis status.

\begin{table}[p]\centering\tiny\renewcommand{\arraystretch}{0.95}\setlength{\tabcolsep}{3pt}
\caption{Item measurement table showing candidate WVS indicators, proposed GPS links, abbreviated questionnaire wording, recode functions $g(v)$, valid ranges, and analysis status.}\label{tab:items}
\begin{tabularx}{\textwidth}{llXccc}\toprule
WVS & GPS link & Abbreviated wording & $g(v)$ & Range & Status \\\midrule
Q57 & GPS trust & Most people can be trusted (1) vs need to be very careful (2) & $1\{v=1\}$ & 0--1 & primary rectangle \\
Q58 & GPS trust & Trust: your family (1 trust completely \dots\ 4 do not trust at all) & $5-v$ & 1--4 & primary rectangle \\
Q59 & GPS trust & Trust: your neighborhood (1 trust completely \dots\ 4 do not trust at all) & $5-v$ & 1--4 & primary rectangle \\
Q60 & GPS trust & Trust: people you know personally (1 trust completely \dots\ 4 do not trust at all) & $5-v$ & 1--4 & primary rectangle \\
Q61 & GPS trust & Trust: people you meet for the first time (1 trust completely \dots\ 4 do not trust at all) & $5-v$ & 1--4 & primary rectangle \\
Q62 & GPS trust & Trust: people of another religion (1 trust completely \dots\ 4 do not trust at all) & $5-v$ & 1--4 & primary rectangle \\
Q63 & GPS trust & Trust: people of another nationality (1 trust completely \dots\ 4 do not trust at all) & $5-v$ & 1--4 & primary rectangle \\
Q64 & GPS trust & Confidence: the churches (1 a great deal \dots\ 4 none at all). Questionnaire prints `The [churches]1' with footnote: replace with the relevant cultural term. & $5-v$ & 1--4 & primary rectangle \\
Q73 & GPS trust & Confidence: parliament (1 a great deal \dots\ 4 none at all) & $5-v$ & 1--4 & primary rectangle \\
Q43 & GPS patience & Less importance placed on work in our lives: good (1) / don't mind (2) / bad (3) & $v$ & 1--3 & exploratory bridge \\
Q50 & GPS patience & Satisfaction with the financial situation of the household (1 completely dissatisfied \dots\ 10 completely satisfied) & $v$ & 1--10 & exploratory bridge \\
Q106 & GPS risk taking & Incomes should be made more equal (1) vs greater incentives for individual effort (10) & $v$ & 1--10 & exploratory bridge \\
Q107 & GPS risk taking & Private ownership of business and industry should be increased (1) vs government ownership of business and industry should be increased (10) & $v$ & 1--10 & exploratory bridge \\
Q109 & GPS risk taking & Competition is good (1) vs competition is harmful (10) & $v$ & 1--10 & exploratory bridge \\
Q178 & GPS risk taking & Justifiable: avoiding a fare on public transport (1 never justifiable \dots\ 10 always justifiable) & $v$ & 1--10 & exploratory bridge \\
Q81 & GPS positive reciprocity & Confidence: charitable or humanitarian organizations (1 a great deal \dots\ 4 none at all) & $5-v$ & 1--4 & exploratory bridge \\
Q176 & GPS negative reciprocity & Agree/disagree: nowadays one often has trouble deciding which moral rules are the right ones to follow (1 completely agree \dots\ 10 completely disagree) & $v$ & 1--10 & exploratory bridge \\
Q177 & GPS negative reciprocity & Justifiable: claiming government benefits to which you are not entitled (1 never \dots\ 10 always) & $11-v$ & 1--10 & exploratory bridge \\
Q179 & GPS negative reciprocity & Justifiable: stealing property (1 never \dots\ 10 always) & $11-v$ & 1--10 & exploratory bridge \\
Q195 & GPS negative reciprocity & Justifiable: death penalty (1 never \dots\ 10 always) & $v$ & 1--10 & exploratory bridge \\
Q99 & GPS altruism & Membership: environmental organization (2 active member / 1 inactive member / 0 don't belong) & $v$ & 0--2 & exploratory bridge \\
Q101 & GPS altruism & Membership: humanitarian or charitable organization (2 active / 1 inactive / 0 don't belong or not a member) & $v$ & 0--2 & exploratory bridge \\
Q103 & GPS altruism & Membership: self-help group, mutual aid group (2 active / 1 inactive / 0 don't belong) & $v$ & 0--2 & exploratory bridge \\
Q69 & GPS trust & Confidence: the police (1 a great deal \dots\ 4 none at all) & $5-v$ & 1--4 & sensitivity only \\
Q70 & GPS trust & Confidence: the courts (1 a great deal \dots\ 4 none at all) & $5-v$ & 1--4 & sensitivity only \\
Q71 & GPS trust & Confidence: the government (1 a great deal \dots\ 4 none at all) & $5-v$ & 1--4 & sensitivity only \\
\bottomrule
\end{tabularx}
\par\vspace{3pt}
\raggedright\emph{Note:} 26 item rows from the measurement specification: 23 in the primary rectangle and Q69--Q71 as sensitivity-only rows outside that rectangle. Patience Q43 and Q50 are separate rows rather than a pooled composite. Wording, recodes $g(v)$, ranges, and status follow the project measurement inventory.
\end{table}

The main intervals use common country-bootstrap draws (seed 20260905; 2,000 draws), average ranks for ties, and undefined values for constant vectors. Sign-profile resampling retains every member of each selected profile; the equal-profile analysis instead averages within profiles before ranking. Leave-country-out and leave-item-out results describe influence on this observed panel. None of these intervals includes repeated model fitting, respondent resampling, or GPS measurement error. Retrained anchor-permutation experiments, new scenario banks, repeated training seeds, and independent survey batteries remain unrun validation tasks.

The same recodes on the forty-two-country GPS--WVS intersection supply a human-only criterion check on the map (Table~\ref{tab:human42}). Adapter scores are not recomputed in that table.

In Table~\ref{tab:human42}, $n$ denotes the number of countries entering the row.

\input{tables/human42}

\section{Proposed validation design}\label{app:humanscreen}
This manuscript does not field a new survey. The intended use of the fitted policies is to score candidate item text against declared GPS anchors before a human pilot. The following protocol is recorded so that a later study can test that use without changing the present construction.

A researcher predeclares a battery of candidate items, their recodes, and which GPS coordinate each item is proposed to track. Frozen adapters from the present study score those items with the space-prefixed code-likelihood readout without country tokens. Country composites are then ranked against GPS $z$. Interpretation is comparative. Items that transfer the coarse GPS sign about as strongly as the nine-item trust map does in the forty-two-country human panel are retained as aligned candidates. Items that transfer the coarse sign where the corresponding human association on already-fielded items is near zero are set aside as likely echo of the training signs. Items that fail to transfer the coarse sign where the human map is strong are set aside as non-transporting.

A human check then fields a small number of retained and flagged items in a predeclared set of countries, using the same recodes. The confirmatory estimand is human--GPS country-rank association on the new items, with a secondary adapter--human association. The study does not treat adapter--GPS alone as construct validation. Ethics review, consent, and sample-size planning belong to that later study.

%% file: tables/human42.tex
\begin{table}[htbp]\centering\small
\caption{Human--GPS country-rank associations on the sixteen-country item map, using the same recodes as the sixteen-country rectangle. Rows require at least fifty valid weighted responses per item. Complete-item rows retain only countries observed on every listed item. Intervals are 95\% country bootstrap (2,000 draws) on the human panel and omit GPS measurement error.}\label{tab:human42}
\begin{tabular}{lccc}\toprule
Surface & $n$ & $\rho$ & Interval\\\midrule
9 trust items (complete) & 42 & 0.35 & [0.09, 0.56] \\
Q69--Q71 only (complete) & 41 & 0.45 & [0.15, 0.67] \\
12 trust items including Q69--Q71 & 41 & 0.43 & [0.14, 0.64] \\
Q43 vs GPS patience & 42 & $-0.53$ & [$-0.74$, $-0.27$] \\
Q50 vs GPS patience & 42 & 0.39 & [0.09, 0.63] \\
Risk-taking map (complete) & 41 & $-0.06$ & [$-0.43$, 0.31] \\
Positive reciprocity Q81 & 42 & 0.05 & [$-0.27$, 0.39] \\
Negative reciprocity map & 42 & 0.42 & [0.17, 0.59] \\
Altruism map (complete) & 42 & 0.03 & [$-0.31$, 0.35] \\
\bottomrule
\end{tabular}
\end{table}

%% file: references.bib
@article{argyle2023,
  author  = {Argyle, Lisa P. and Busby, Ethan C. and Fulda, Nancy and Gubler, Joshua R. and Rytting, Christopher and Wingate, David},
  title   = {Out of One, Many: Using Language Models to Simulate Human Samples},
  journal = {Political Analysis},
  volume  = {31},
  number  = {3},
  pages   = {337--351},
  year    = {2023},
  doi     = {10.1017/pan.2023.2}
}

@article{bisbee2024,
  author  = {Bisbee, James and Clinton, Joshua D. and Dorff, Cassy and Kenkel, Brenton and Larson, Jennifer M.},
  title   = {Synthetic Replacements for Human Survey Data? The Perils of Large Language Models},
  journal = {Political Analysis},
  volume  = {32},
  number  = {4},
  pages   = {401--416},
  year    = {2024},
  doi     = {10.1017/pan.2024.5}
}

@misc{buskirk2026,
      title={More Parameters Than Populations: A Systematic Literature Review of Large Language Models within Survey Research}, 
      author={Trent D. Buskirk and Florian Keusch and Leah von der Heyde and Adam Eck},
      year={2026},
      eprint={2509.03391},
      archivePrefix={arXiv},
      primaryClass={cs.DL},
      url={https://arxiv.org/abs/2509.03391}, 
}

@article{boelaert2025,
  author  = {Boelaert, Julien and Coavoux, Samuel and Ollion, {\'E}tienne and Petev, Ivaylo and Pr{\"a}g, Patrick},
  title   = {Machine Bias. How Do Generative Language Models Answer Opinion Polls?},
  journal = {Sociological Methods \& Research},
  volume  = {54},
  number  = {3},
  pages   = {1156--1196},
  year    = {2025},
  doi     = {10.1177/00491241251330582}
}

@inproceedings{cao2025,
  author    = {Cao, Yong and Liu, Haijiang and Arora, Arnav and Augenstein, Isabelle and R{\"o}ttger, Paul and Hershcovich, Daniel},
  title     = {Specializing Large Language Models to Simulate Survey Response Distributions for Global Populations},
  booktitle = {Proceedings of the 2025 Conference of the Nations of the Americas Chapter of the Association for Computational Linguistics: Human Language Technologies (Volume 1: Long Papers)},
  pages     = {3141--3154},
  year      = {2025},
  address   = {Albuquerque, New Mexico},
  publisher = {Association for Computational Linguistics},
  doi       = {10.18653/v1/2025.naacl-long.162}
}

@misc{capra2025,
  author        = {Capra, C. M{\'o}nica and Gonzalez-Bonorino, Augusto and Pantoja, Emilio},
  title         = {{LLMs} Can Model Non-{WEIRD} Populations: Experiments with Synthetic Cultural Agents},
  year          = {2025},
  eprint        = {2501.06834},
  archivePrefix = {arXiv},
  primaryClass  = {cs.AI},
  doi           = {10.48550/arXiv.2501.06834}
}

@article{campbellfiske1959,
  author  = {Campbell, Donald T. and Fiske, Donald W.},
  title   = {Convergent and Discriminant Validation by the Multitrait-Multimethod Matrix},
  journal = {Psychological Bulletin},
  volume  = {56},
  number  = {2},
  pages   = {81--105},
  year    = {1959},
  doi     = {10.1037/h0046016}
}

@unpublished{dalloul2026,
  author = {Dalloul, Ami and Pfeifer, Moritz},
  title  = {Can {LLMs} Mimic Household Surveys? From Representative Agents to Population Distributions},
  year   = {2026},
  note   = {{SSRN} 6703798},
  doi    = {10.2139/ssrn.6703798}
}

@article{davidov2014,
  author  = {Davidov, Eldad and Meuleman, Bart and Cieciuch, Jan and Schmidt, Peter and Billiet, Jaak},
  title   = {Measurement Equivalence in Cross-National Research},
  journal = {Annual Review of Sociology},
  volume  = {40},
  pages   = {55--75},
  year    = {2014},
  doi     = {10.1146/annurev-soc-071913-043137}
}

@inproceedings{dettmers2023,
  author        = {Dettmers, Tim and Pagnoni, Artidoro and Holtzman, Ari and Zettlemoyer, Luke},
  title         = {{QL}o{RA}: Efficient Finetuning of Quantized {LLMs}},
  booktitle     = {Advances in Neural Information Processing Systems},
  volume        = {36},
  year          = {2023},
  eprint        = {2305.14314},
  archivePrefix = {arXiv},
  primaryClass  = {cs.LG}
}

@article{falk2018,
  author  = {Falk, Armin and Becker, Anke and Dohmen, Thomas and Enke, Benjamin and Huffman, David and Sunde, Uwe},
  title   = {Global Evidence on Economic Preferences},
  journal = {The Quarterly Journal of Economics},
  volume  = {133},
  number  = {4},
  pages   = {1645--1692},
  year    = {2018},
  doi     = {10.1093/qje/qjy013}
}

@techreport{horton2023,
  author      = {Horton, John J. and Filippas, Apostolos and Manning, Benjamin S.},
  title       = {Large Language Models as Simulated Economic Agents: What Can We Learn from Homo Silicus?},
  type        = {Working Paper},
  number      = {31122},
  institution = {National Bureau of Economic Research},
  year        = {2023},
  month       = apr,
  note        = {Revised February 2026},
  doi         = {10.3386/w31122}
}

@inproceedings{hu2022,
  author        = {Hu, Edward J. and Shen, Yelong and Wallis, Phillip and Allen-Zhu, Zeyuan and Li, Yuanzhi and Wang, Shean and Wang, Lu and Chen, Weizhu},
  title         = {{L}o{RA}: Low-Rank Adaptation of Large Language Models},
  booktitle     = {International Conference on Learning Representations},
  year          = {2022},
  eprint        = {2106.09685},
  archivePrefix = {arXiv},
  primaryClass  = {cs.CL}
}

@inproceedings{krsteski2026,
  author    = {Krsteski, Stefan and Russo, Giuseppe and Chang, Serina and West, Robert and Gligori{\'c}, Kristina},
  title     = {Valid Survey Simulations with Limited Human Data: The Roles of Prompting, Fine-Tuning, and Rectification},
  booktitle = {Proceedings of the 64th Annual Meeting of the Association for Computational Linguistics},
  pages     = {10887--10906},
  year      = {2026},
  publisher = {Association for Computational Linguistics},
  doi       = {10.18653/v1/2026.acl-long.498}
}

@article{bradleyterry1952,
 ISSN = {00063444, 14643510},
 URL = {http://www.jstor.org/stable/2334029},
 author = {Ralph Allan Bradley and Milton E. Terry},
 journal = {Biometrika},
 number = {3/4},
 pages = {324--345},
 publisher = {[Oxford University Press, Biometrika Trust]},
 title = {Rank Analysis of Incomplete Block Designs: I. The Method of Paired Comparisons},
 urldate = {2026-09-05},
 volume = {39},
 year = {1952}
}

@article{ludwig2026,
  author  = {Ludwig, Jens and Mullainathan, Sendhil and Rambachan, Ashesh},
  title   = {Large Language Models: An Applied Econometric Framework},
  journal = {Annual Review of Economics},
  volume  = {18},
  pages   = {283--316},
  year    = {2026},
  doi     = {10.1146/annurev-economics-120925-105620}
}

@inproceedings{rafailov2023,
  author        = {Rafailov, Rafael and Sharma, Archit and Mitchell, Eric and Ermon, Stefano and Manning, Christopher D. and Finn, Chelsea},
  title         = {Direct Preference Optimization: Your Language Model Is Secretly a Reward Model},
  booktitle     = {Advances in Neural Information Processing Systems},
  volume        = {36},
  year          = {2023},
  eprint        = {2305.18290},
  archivePrefix = {arXiv},
  primaryClass  = {cs.LG}
}

@article{steenkamp1998,
  author  = {Steenkamp, Jan-Benedict E. M. and Baumgartner, Hans},
  title   = {Assessing Measurement Invariance in Cross-National Consumer Research},
  journal = {Journal of Consumer Research},
  volume  = {25},
  number  = {1},
  pages   = {78--107},
  year    = {1998},
  doi     = {10.1086/209528}
}

@inproceedings{suh2025,
  author    = {Suh, Joseph and Jahanparast, Erfan and Moon, Suhong and Kang, Minwoo and Chang, Serina},
  title     = {Language Model Fine-Tuning on Scaled Survey Data for Predicting Distributions of Public Opinions},
  booktitle = {Proceedings of the 63rd Annual Meeting of the Association for Computational Linguistics (Volume 1: Long Papers)},
  pages     = {21147--21170},
  year      = {2025},
  publisher = {Association for Computational Linguistics},
  doi       = {10.18653/v1/2025.acl-long.1028}
}

@article{wang2025,
  author  = {Wang, Angelina and Morgenstern, Jamie and Dickerson, John P.},
  title   = {Large Language Models That Replace Human Participants Can Harmfully Misportray and Flatten Identity Groups},
  journal = {Nature Machine Intelligence},
  volume  = {7},
  pages   = {400--411},
  year    = {2025},
  doi     = {10.1038/s42256-025-00986-z}
}

@misc{wvs2022,
  author       = {Haerpfer, Christian and Inglehart, Ronald and Moreno, Alejandro and Welzel, Christian and Kizilova, Kseniya and Diez-Medrano, Jaime and Lagos, Marta and Norris, Pippa and Ponarin, Eduard and Puranen, Bi and others},
  title        = {World Values Survey Wave 7 (2017--2022) Cross-National Data-Set},
  year         = {2022},
  howpublished = {Madrid, Spain and Vienna, Austria: JD Systems Institute and WVSA Secretariat},
  note         = {Version 4.0.0},
  doi          = {10.14281/18241.18}
}

@inproceedings{li2024culture,
 author={Li, Cheng and Chen, Mengzhuo and Wang, Jindong and Sitaram, Sunayana and Xie, Xing},
 title={{CultureLLM}: Incorporating Cultural Differences into Large Language Models},
 booktitle={Advances in Neural Information Processing Systems}, volume={37},year={2024},
 url={https://proceedings.neurips.cc/paper_files/paper/2024/hash/9a16935bf54c4af233e25d998b7f4a2c-Abstract-Conference.html}}

@misc{jung2026,
 author={Jung, MinJae and Kim, Minwoo},
 title={Korean Culture into {LLM} Alignment: Toward Cultural Coherence},
 year={2026},eprint={2606.06797},archivePrefix={arXiv},url={https://arxiv.org/abs/2606.06797}}
